\PassOptionsToPackage{unicode}{hyperref}
\PassOptionsToPackage{hyphens}{url}
\PassOptionsToPackage{dvipsnames,svgnames*,x11names*}{xcolor}
\documentclass[
  11pt,
]{article}
\usepackage{amsmath,amssymb}
\usepackage{lmodern}
\usepackage{setspace}
\usepackage{iftex}
\ifPDFTeX
  \usepackage[T1]{fontenc}
  \usepackage[utf8]{inputenc}
  \usepackage{textcomp} 
\else 
  \usepackage{unicode-math}
  \defaultfontfeatures{Scale=MatchLowercase}
  \defaultfontfeatures[\rmfamily]{Ligatures=TeX,Scale=1}
\fi
\IfFileExists{upquote.sty}{\usepackage{upquote}}{}
\IfFileExists{microtype.sty}{
  \usepackage[]{microtype}
  \UseMicrotypeSet[protrusion]{basicmath} 
}{}
\makeatletter
\@ifundefined{KOMAClassName}{
    \setlength{\parindent}{0pt}
    \setlength{\parskip}{4pt}
}{
  \KOMAoptions{parskip=half}}
\makeatother
\usepackage{xcolor}
\IfFileExists{xurl.sty}{\usepackage{xurl}}{} 
\IfFileExists{bookmark.sty}{\usepackage{bookmark}}{\usepackage{hyperref}}
\hypersetup{
  pdftitle={GitHub is an effective platform for collaborative and reproducible laboratory research},
  colorlinks=true,
  linkcolor={blue},
  filecolor={Maroon},
  citecolor={Blue},
  urlcolor={blue},
  pdfcreator={LaTeX via pandoc}}
\usepackage[margin=0.6in]{geometry}
\usepackage{longtable,booktabs,array}
\usepackage{calc} 
\usepackage{etoolbox}
\makeatletter
\patchcmd\longtable{\par}{\if@noskipsec\mbox{}\fi\par}{}{}
\makeatother
\usepackage{graphicx}
\makeatletter
\def\maxwidth{\ifdim\Gin@nat@width>\linewidth\linewidth\else\Gin@nat@width\fi}
\def\maxheight{\ifdim\Gin@nat@height>\textheight\textheight\else\Gin@nat@height\fi}
\makeatother
\setkeys{Gin}{width=\maxwidth,height=\maxheight,keepaspectratio}
\makeatletter
\def\fps@figure{H}
\makeatother
\usepackage{lineno}
\usepackage[colaction]{multicol}

\usepackage[font=small,labelfont=bf]{caption}
\usepackage[scaled]{helvet}

\usepackage[T1]{fontenc}
\ifLuaTeX
  \usepackage{selnolig}  
\fi
\newlength{\cslhangindent}
\newlength{\csllabelwidth}
\newenvironment{CSLReferences}[2] 
 {
  \setlength{\parindent}{0pt}
  \ifnum #2 > 0
  \fi
 }%
 {}
\usepackage{calc}

\newcommand{\CSLLeftMargin}[1]{\parbox[t]{\csllabelwidth}{#1}}
\newcommand{\CSLRightInline}[1]{\parbox[t]{\linewidth - \csllabelwidth}{#1}\break}

\title{GitHub is an effective platform for collaborative and
reproducible laboratory research}
\author{}
\date{}

\usepackage{float}
\makeatletter
\renewcommand{\section}{
\@startsection
{section}
{1}
{\z@}
{2pt}
{1pt}
{\bfseries\large}} 
\renewcommand{\subsection}{
\@startsection
{subsection}
{2}
{\z@}
{0pt}
{1pt}
{\bfseries}} 
\renewcommand{\subsubsection}{
\@startsection
{subsubsection}
{3}
{\z@}
{0pt}
{1pt}
{\bfseries}} 
\renewcommand{\paragraph}{
\@startsection
{paragraph}
{4}
{\z@}
{0pt}
{1pt}
{\itshape\bfseries}} 
\renewcommand{\subparagraph}{
\@startsection
{subparagraph}
{5}
{\z@}
{0pt}
{1pt}
{\bfseries\itshape}} 

\makeatletter
\renewcommand{\maketitle}{
\begin{flushleft}
  \Large\textbf{\@title}\vskip6pt
\end{flushleft}
}

\begin{document}
\maketitle

\setstretch{1}
Katharine Y. Chen\textsuperscript{1}, Maria
Toro-Moreno\textsuperscript{1,†}
\href{https://orcid.org/0000-0002-0497-2259}{\includegraphics[width=0.4cm,height=\textheight]{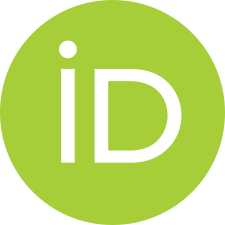}},
Arvind Rasi Subramaniam\textsuperscript{1,†}
\href{https://orcid.org/0000-0001-6145-4303}{\includegraphics[width=0.4cm,height=\textheight]{./svg/orcid.logo.icon.png}}

\textsuperscript{1} Basic Sciences Division and Computational Biology
Section of the Public Health Sciences Division, Fred Hutchinson Cancer
Center, Seattle, USA

\textsuperscript{†} Corresponding authors: M.T.M:
\href{mailto:mtoromor@fredhutch.org}{\nolinkurl{mtoromor@fredhutch.org}},
A.R.S: \href{mailto:rasi@fredhutch.org}{\nolinkurl{rasi@fredhutch.org}}

\hypertarget{abstract}{%
\subsection{Abstract}\label{abstract}}

Laboratory research is a complex, collaborative process that involves
several stages, including hypothesis formulation, experimental design,
data generation and analysis, and manuscript writing. Although
reproducibility and data sharing are increasingly prioritized at the
publication stage, integrating these principles at earlier stages of
laboratory research has been hampered by the lack of broadly applicable
solutions. Here, we propose that the workflow used in modern software
development offers a robust framework for enhancing reproducibility and
collaboration in laboratory research. In particular, we show that
GitHub, a platform widely used for collaborative software projects, can
be effectively adapted to organize and document all aspects of a
research project's lifecycle in a molecular biology laboratory. We
outline a three-step approach for incorporating the GitHub ecosystem
into laboratory research workflows: 1. designing and organizing
experiments using issues and project boards, 2. documenting experiments
and data analyses with a version control system, and 3. ensuring
reproducible software environments for data analyses and writing tasks
with containerized packages. The versatility, scalability, and
affordability of this approach make it suitable for various scenarios,
ranging from small research groups to large, cross-institutional
collaborations. Adopting this framework from a project's outset can
increase the efficiency and fidelity of knowledge transfer within and
across research laboratories. An example GitHub repository based on this
approach is available at \url{https://github.com/rasilab/github_demo}
and a template repository that can be copied is available at
\url{https://github.com/rasilab/github_template}.

\hypertarget{introduction}{%
\subsection{Introduction}\label{introduction}}

Scientific progress is contingent on the ability to reproduce and build
upon previous findings. To promote reproducibility of published studies,
journals and funding agencies increasingly require researchers to make
their data and analysis methods available in public
repositories\textsuperscript{\protect\hyperlink{ref-Hrynaszkiewicz2020}{1}--\protect\hyperlink{ref-nih2023}{3}}.
In parallel, data-intensive fields such as machine learning,
computational biology, and ecology have developed workflows and tools
that facilitate computational
reproducibility\textsuperscript{\protect\hyperlink{ref-Heil2021}{4}--\protect\hyperlink{ref-Jenkins2023}{7}}.
Reproducibility is also a critical element of collaborative research,
where multiple researchers need to share and build upon each other's
work. Indeed, many reproducibility standards and tools have been
developed in the context of large-scale, cross-institutional
collaborations\textsuperscript{\protect\hyperlink{ref-Diaba-Nuhoho2021}{8}--\protect\hyperlink{ref-Braga2023}{12}}.

Compared to computational research, reproducibility and collaboration
are less emphasized during the course of laboratory research, especially
in the context of small research groups. This is perhaps because
projects within a laboratory are often executed by individuals, and are
rarely framed as collaborative efforts. However, viewed from a broader
perspective, collaboration is an integral and ubiquitous feature of even
small laboratory research groups. Group members discuss ideas, exchange
protocols, and share reagents and data on a daily basis. Trainees
collaborate with group leaders in hypothesis formulation, data
interpretation, literature review, and manuscript writing. Most
importantly, individuals ``collaborate'' with their future selves by
analyzing and building on their own results, and also with future lab
members who extend their work after they have left the lab. Many
publicized instances of post-publication irreproducibility arise from
the inability of either the same scientist at a later time or a
different scientist within the same laboratory to replicate previous
results\textsuperscript{\protect\hyperlink{ref-Berg2017}{13},\protect\hyperlink{ref-Wosen2023}{14}}.
Thus, workflows that improve scientific documentation and collaboration
within laboratory groups from the outset of a project can enhance
reproducibility of published studies across the wider research
community.

While much effort has been devoted to improving reproducibility during
data analysis and post-publication stages of research, these do not
address earlier critical stages of laboratory research. Even in a `wet'
lab where experiments are physically performed, researchers spend
significant time and resources on literature review, hypothesis
formulation, experimental design, in addition to the generation,
visualization and interpretation of data. Nevertheless, few tools and
workflows exist to document and share these activities in a structured
and reproducible manner. Lab notebooks are the most common media used to
document laboratory research, but they are typically only used for
recording methods and data. An NIH handbook on lab notebooks, for
example, explicitly discourages the inclusion of speculative ideas or
informal
conversations\textsuperscript{\protect\hyperlink{ref-Ryan2010}{15}},
despite the recognition that these are often the source of scientific
breakthroughs. Electronic lab notebooks (ELNs), despite their
popularity, are stored in proprietary formats, incur a recurrent cost,
tend to become defunct over time, and have poor interoperability with
each other\textsuperscript{\protect\hyperlink{ref-Higgins2022}{16}}.
Cloud-based tools like Google Docs, Dropbox, and Sharepoint allow
sharing of data and documents, but do not provide a structured way to
track changes over time or record project-related communication. Email
and messaging tools such as Slack and Microsoft Teams facilitate
informal discussion of ideas and data, but these are poorly suited for
organizing data and discussion in a reproducible manner. Thus, it is
increasingly common for data, experiment details, analysis code, and
project communication to be fragmented across multiple tools within
laboratory research groups.

\begin{figure}
\centering
\includegraphics[width=5.87in,height=\textheight]{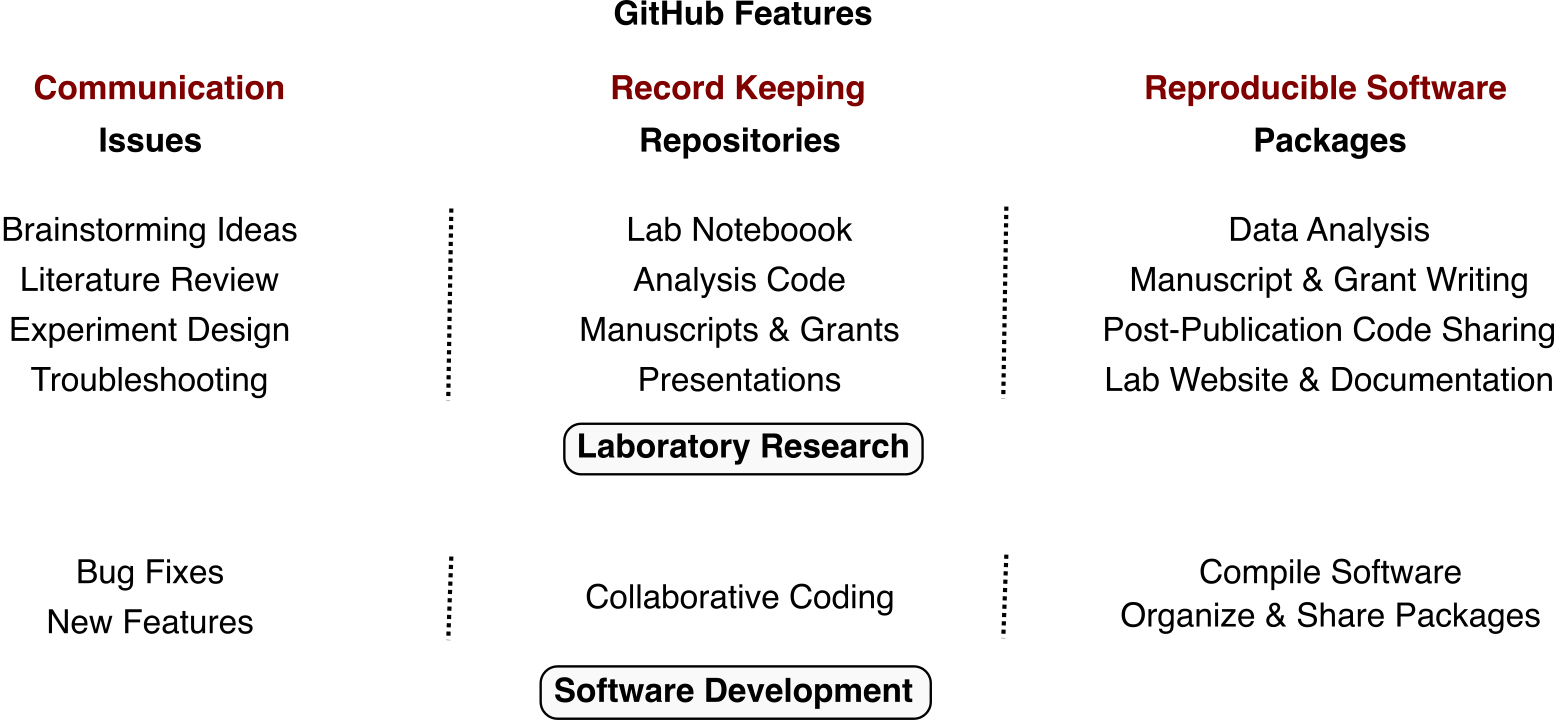}
\caption{\label{fig:tools}}
\end{figure}

The process of software development bears several similarities to
activities in laboratory research (Figure \ref{fig:tools}). Developing
software involves understanding the state-of-the-art solutions to a
problem, formulating a hypothesis for improving existing approaches,
breaking down the hypothesis into smaller testable features, writing
code to implement features, analyzing the outcome of each feature
development, and troubleshooting and fixing errors when necessary.
Software development occurs in groups of all sizes ranging from a single
developer to thousands of contributors, often distributed across
different timezones, working on a common project. The necessity to
document and share all stages of software development between
contributors has led to the emergence of mature tools and workflows that
facilitate reproducibility and collaboration. These include concepts
such as issue
tracking\textsuperscript{\protect\hyperlink{ref-Johnson2003}{17}},
version
control\textsuperscript{\protect\hyperlink{ref-Blischak2016}{18}}, and
containerization\textsuperscript{\protect\hyperlink{ref-Moreau2023}{19}}
that have become integral to most software development projects.

Many of the common workflows associated with software development are
implemented in GitHub, a popular cloud-based platform used by over 100
million developers worldwide and 90\% of Fortune 100
companies\textsuperscript{\protect\hyperlink{ref-2023}{20},\protect\hyperlink{ref-stackoverflow2023}{21}}.
GitHub repositories are highly scalable, with both small projects built
by single developers and large projects with over 2,000
contributors\textsuperscript{\protect\hyperlink{ref-octoverse2022}{22}}.
In the scientific community, GitHub is used to share data analysis
workflows after
publication\textsuperscript{\protect\hyperlink{ref-Cadwallader2022}{23},\protect\hyperlink{ref-Perkel2016}{24}},
develop and share computational
tools\textsuperscript{\protect\hyperlink{ref-Perez-Riverol2016}{25}},
perform individual record
keeping\textsuperscript{\protect\hyperlink{ref-Ram2013}{10},\protect\hyperlink{ref-Stanisic2015}{26},\protect\hyperlink{ref-Chure2022}{27}},
conduct open science
projects\textsuperscript{\protect\hyperlink{ref-Lowndes2017}{11},\protect\hyperlink{ref-Braga2023}{12}},
and collaboratively write
manuscripts\textsuperscript{\protect\hyperlink{ref-Ram2013}{10},\protect\hyperlink{ref-Himmelstein2019}{28}}.
However, how the standard workflows and rich features of GitHub (Figure
\ref{fig:tools}) can be adapted to improve reproducibility and
collaboration within a traditional laboratory research group has not
been explored.

\begin{figure}
\centering
\includegraphics[width=5in,height=\textheight]{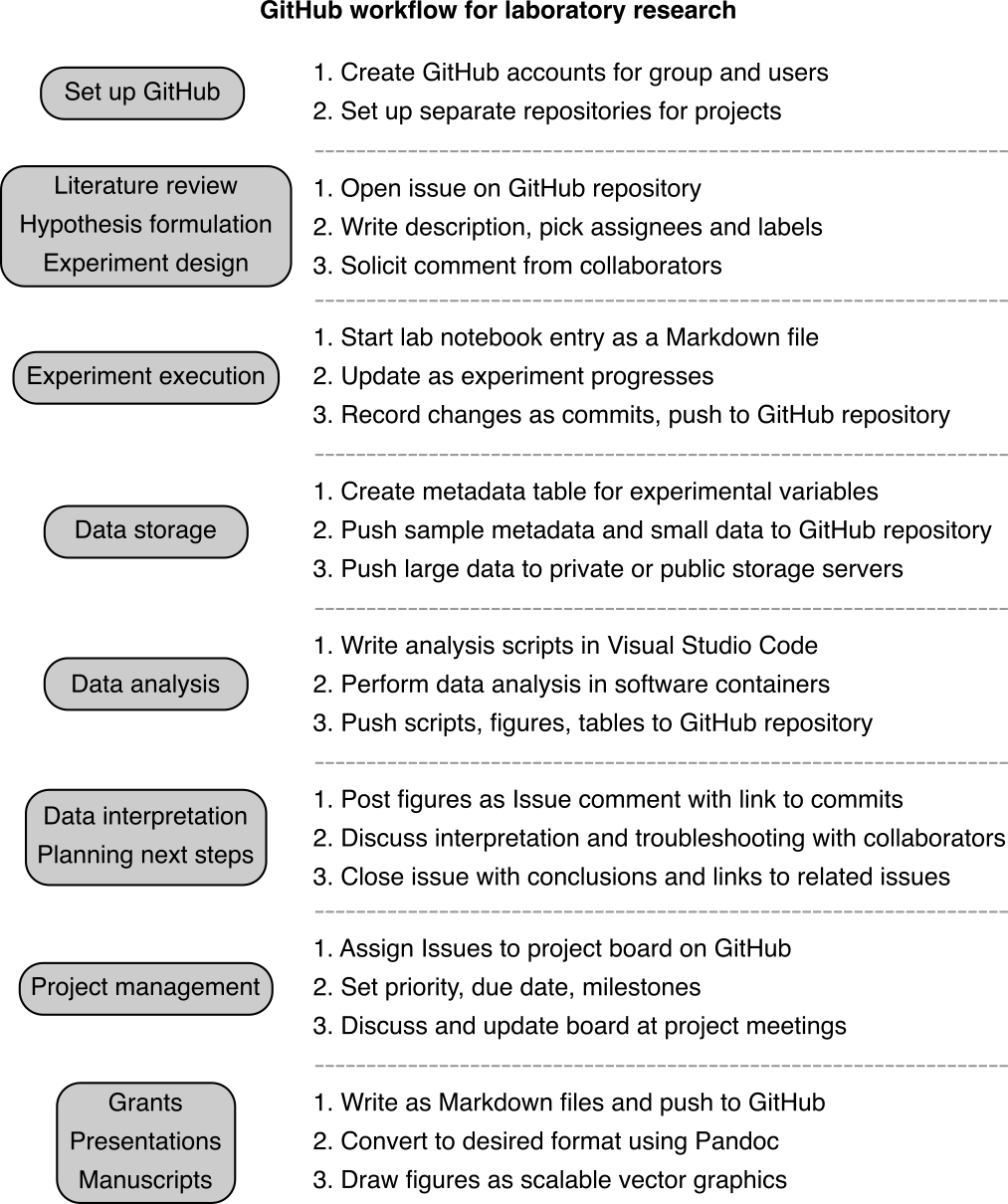}
\caption{\label{fig:workflow}}
\end{figure}

Here, we aim to provide a practical demonstration of GitHub's use in the
context of a laboratory research group centered around molecular biology
experiments. Our goal is to show how GitHub provides an intuitive and
structured framework to organize and document all aspects of a research
project's lifecycle, including literature review, hypothesis
formulation, experiment design, lab work, data analysis, and manuscript
and grant writing (Figure \ref{fig:workflow}). Like many academic
researchers, we started using GitHub for version control of data
analysis scripts. Over the past nine years, we have expanded our use of
GitHub to document all aspects of research in our laboratory, and manage
collaborations both within our institution and with external
collaborators. None of us had any formal training in software
development or the use of GitHub. We arrived at our current workflow
through trial and error, tutorials on the internet, and by seeking help
from more experienced users within and outside our lab. We find that
starting a project with this framework enables us to take full advantage
of GitHub's many features, though adopting this workflow at any stage of
a project can still be beneficial.

\hypertarget{set-up-github-for-laboratory-research}{%
\subsection{Set up GitHub for laboratory
research}\label{set-up-github-for-laboratory-research}}

\begin{figure}
\centering
\includegraphics[width=5.5in,height=\textheight]{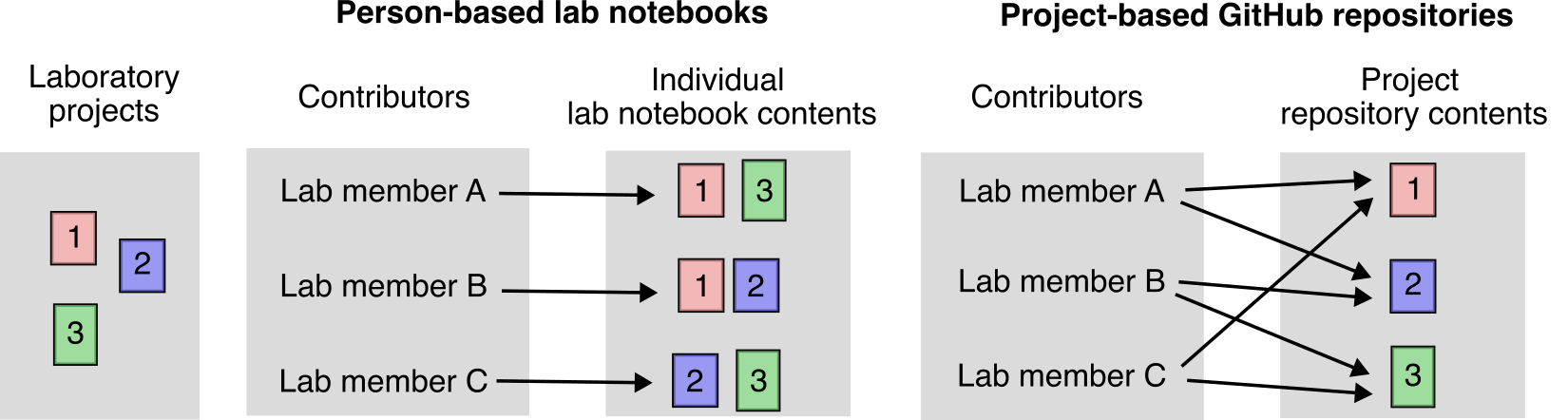}
\caption{\label{fig:person_vs_project}}
\end{figure}

GitHub is centered around the concept of repositories, which can be
thought of as cloud-based folders that contain all files and
documentation related to a specific research project. While a single
GitHub repository can be used to record all work by a single user, akin
to a traditional lab notebook, organizing repositories based on projects
provides better structure for collaborative research (Figure
\ref{fig:person_vs_project}). Repositories have a unique name, which is
often the project name, and a standardized URL in the format
https://github.com/GROUP\_NAME/PROJECT\_NAME (for example,
https://github.com/rasilab/ribosome\_collisions\_yeast). Each GitHub
repository's content can also be edited on a local computer, allowing
users to work offline and synchronize with the cloud when they connect
to the internet. To edit repository files locally, we typically use
Visual Studio Code, a popular open source editor that works seamlessly
with GitHub and has extensive features for writing and data analysis.

Adopting a GitHub-based workflow within a group or a team starts with a
designated administrator creating a GitHub organization. Then, each
member of the group creates a GitHub account for themselves, and are
made members of the GitHub organization by the administrator. Once the
organization and user accounts are set up, the administrator or any
group member can create a new repository for each project the group is
working on. All work and communication related to each project will be
recorded within the corresponding repository. A practically unlimited
number of repositories can be created within a GitHub organization, and
access to each repository can be controlled by the administrator. All
functionalities that we describe in the following sections are currently
available as part of the free GitHub plan. Research groups in
educational and non-profit institutions also get free access to the
GitHub Team plan, which can be useful for accessing more advanced
features.

\hypertarget{use-issues-to-organize-and-collaborate}{%
\subsection{Use issues to organize and
collaborate}\label{use-issues-to-organize-and-collaborate}}

Experiments are the fundamental units of laboratory research projects.
Yet few standards exist to conduct the design, execution, documentation,
and interpretation of experiments in a structured, reproducible, and
collaborative manner. Even what constitutes a single experiment is often
unclear from perusing lab notebooks. Notebooks are typically
chronological records of all activities by a single researcher, making
it difficult to isolate individual experiments. Further, the fragmented
tools for laboratory research are reflected in the fragmented nature of
the experiments themselves. Design and execution steps are recorded in
lab notebooks, physical samples are stored in freezers or shelves, data
is kept in centralized servers, analysis scripts are backed up in
personal computers, and ideas, hypotheses, and interpretation are
discussed in person or over electronic messaging apps. Tracking the full
trajectory of an experiment across email discussions, physical samples,
data, and analysis scripts becomes a challenge, particularly for future
lab members who may need to build upon the work after the lead
researcher leaves the project.

The GitHub `issues' feature provides an intuitive and flexible interface
to organize and collaborate on all aspects of a laboratory experiment.
In software projects, issues were originally used to track bugs or
problems (hence the name `issue'), but their utility has expanded to new
feature proposals, maintenance tasks, and general discussion
topics\textsuperscript{\protect\hyperlink{ref-Johnson2003}{17}}.
Similarly, in laboratory research, we use issues not just for
troubleshooting, but for organizing and discussing all aspects of
research, from hypothesis formulation and experiment design to data
interpretation and manuscript writing. Each issue is limited to a single
topic, which focuses the ensuing discussion and resolution. In GitHub,
issues are given a unique number and a URL
(https://github.com/GROUP\_NAME/PROJECT\_NAME/issues/ISSUE\_NUMBER) that
provide a centralized location to track all work and discussion related
to that issue. Each issue has a description field and optional
commenting fields, which can be used to write, attach files, and paste
images.

\begin{figure}
\centering
\includegraphics[width=7.3in,height=\textheight]{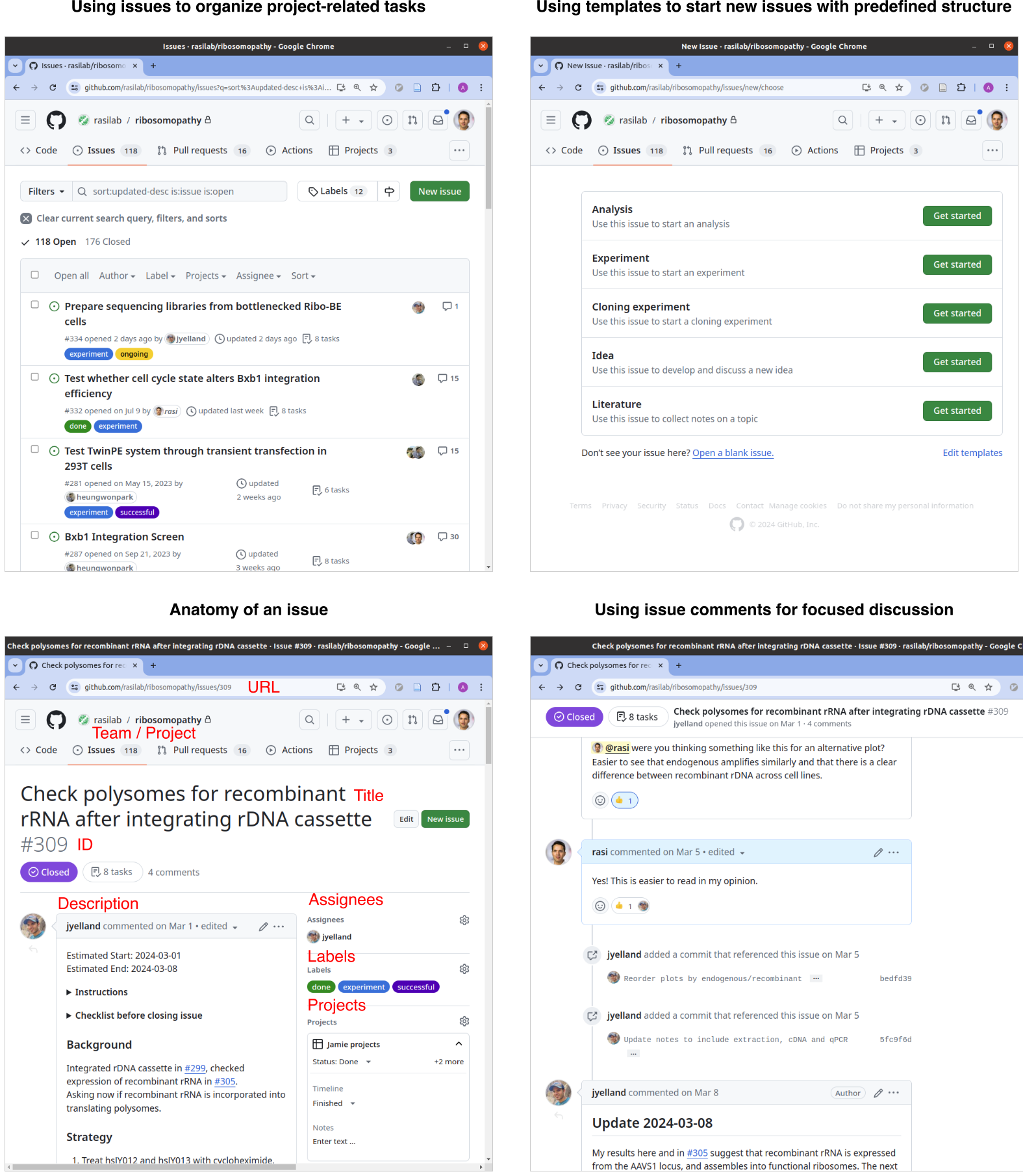}
\caption{\label{fig:issues}}
\end{figure}

In our research group, each experiment begins with the creation of a new
issue in the corresponding project repository by any of the project
members (Figure \ref{fig:issues}). The issue description is used to
describe the rationale and background of the experiment and the strategy
for performing the experiment. Project members can discuss aspects of
experimental design, provide clarification in the comments section, and
update the issue description as needed. Once an experiment is started,
the comments section is used to discuss troubleshooting steps,
intermediate data and figures, and interpretation of results. The issue
number provides a convenient way to reference the experiment across
physical samples, work logs, computer file names, and discussions in
other issues. For instance, we include the issue number as a prefix on
the labels of sample tubes along with suffixes denoting the sample type
or condition, which enables succinct and unambiguous tracking of samples
by all lab members. Once an experiment is completed, the issue
description is updated with key conclusions, tables, and figures, and
the issue is `closed'. Even if the experiment proposed in an issue is
paused or ultimately not pursued, there is a record of the
decision-making process and the issue can be re-opened by a project
member at any time. Finally, we use issues not just for experiments, but
also for discussing broader ideas for projects, reviewing specific
literature topics, and for collaborating on grant proposals and
manuscripts. For complex experiments with multi-stage design and
analysis steps, we create separate issues to document the design and
analysis steps.

GitHub provides a number of features to organize and prioritize issues
within a project and across projects (Figure \ref{fig:issues}). One or
more `assignees' can be associated with each issue to ensure that they
receive notifications about any work or discussion related to the issue,
and to track responsibilities. Color-coded `labels' can be used to
distinguish between different issue types such as `experiment', `data
analysis', `literature review', or `project idea', or to indicate the
state of the issue such as `todo', `ongoing', `paused', `completed',
`abandoned'. Issues can be grouped together into `milestones' to track
progress towards a specific goal or deadline. For example, we use
milestones to group issues that need to be completed for a figure in a
manuscript or a grant. `Issue templates' can also be created to
standardize the format of common issue types across contributors. For
example, an `Experiment' issue template can include prompts to include
relevant background, strategy, and conclusion in the description section
and pre-populate the issue type label and common assignees such as the
lead researcher on th e project. Each time a group member creates a new
issue, they have the option to use one of these issue templates, which
is especially useful for new or junior group members. GitHub `project
boards' provide a higher level visual interface to organize and
prioritize issues across projects and repositories. For example, a group
member can create a project board for themselves and add their own
fields such as due date or priority to each issue. During project
meetings, the project board can be used to quickly understand each group
member's priorities and deadlines, helping ensure that all collaborators
are on the same page.

In summary, issues, a widely used feature in software development, also
provide an intuitive structure to organize and collaborate across every
stage of laboratory-based projects from hypothesis formulation to
manuscript writing. An issue-based workflow enables group members to
access and contribute to all project-related documentation and
communication, regardless of the stage in which they join the project.
This can be particularly useful for new lab members, who can quickly get
up to speed on a project by reading through the issue descriptions and
comments. Closed issues can be reopened if needed, which can be useful
for revisiting old experiments or ideas. Thus, by providing a
centralized location for tracking information relevant to a specific
experiment, analysis, or idea, GitHub issues facilitate reproducibility
and knowledge transfer during all stages of laboratory research
projects.

\begin{longtable}[]{@{}
  >{\raggedright\arraybackslash}p{(\columnwidth - 2\tabcolsep) * \real{0.2879}}
  >{\raggedright\arraybackslash}p{(\columnwidth - 2\tabcolsep) * \real{0.7121}}@{}}
\caption{Guides to resources mentioned in this paper}\tabularnewline
\toprule
\begin{minipage}[b]{\linewidth}\raggedright
Resource
\end{minipage} & \begin{minipage}[b]{\linewidth}\raggedright
URL
\end{minipage} \\
\midrule
\endfirsthead
\toprule
\begin{minipage}[b]{\linewidth}\raggedright
Resource
\end{minipage} & \begin{minipage}[b]{\linewidth}\raggedright
URL
\end{minipage} \\
\midrule
\endhead
GitHub & \url{https://docs.github.com/get-started} \\
Markdown & \url{https://www.markdownguide.org/getting-started} \\
Visual Studio Code & \url{https://code.visualstudio.com/docs} \\
Docker & \url{https://docs.docker.com} \\
Git & \url{https://swcarpentry.github.io/git-novice/} \\
Pandoc & \url{https://pandoc.org/MANUAL.html} \\
Semantic versioning & \url{https://semver.org} \\
\bottomrule
\end{longtable}

\hypertarget{use-git-to-store-and-track-your-work}{%
\subsection{Use Git to store and track your
work}\label{use-git-to-store-and-track-your-work}}

When a researcher performs a specific experiment or data analysis task,
they record the execution steps and results along the way. This record
is often maintained in physical or electronic lab notebooks for wet lab
experiments and as code files in computer folders for data analyses.
Once a set of experiments or analyses are completed, the researcher may
write a manuscript or grant proposal that summarizes the results and
conclusions. Such manuscripts or grant proposals are typically written
using software such as Microsoft Word or Google Docs, often in a
collaborative manner with multiple authors. Since these steps can extend
over several months or even years, it is frequently challenging to
maintain an organized and chronological record of the contributions made
by each group member, and to track the changes made to multiple files in
different folders. It is common to have many copies of the same file
with cryptic names like `manuscript\_v3.docx', `annotations\_ARS.xls' to
indicate their provenance. While electronic lab notebooks and `track
change' features in word processing software can help maintain a record
of changes to single files, these tools are neither designed to work
across files of various types, nor to easily identify each author of
overlapping changes.

Git is a version control system that records the history of file
additions and modifications in a folder, and is used by over 90\% of
programmers worldwide to track changes to their
code\textsuperscript{\protect\hyperlink{ref-stackoverflow2023}{21}}. Git
allows multiple copies of the folder to be asynchronously edited across
computers, and GitHub repositories are essentially remote copies of a
local folder tracked using Git. Anyone with access to a GitHub
repository can download a local copy of the folder (`clone' in Git
terminology) , add or edit file in the folder, choose which files they
want to track (`stage'), create a snapshot of the changes (`commit'),
and synchronize with the GitHub repository (`push'). These Git features
are tightly integrated into popular text editors like Visual Studio
Code, which allows users to make, stage, commit, and push changes to a
GitHub repository without leaving the text editor. Each commit is
accompanied by a `commit message' which is a short description of what
changes were made since the previous commit. Importantly, the commit
history of a project serves as an audit trail, recording who did what,
and when.

\begin{figure}
\centering
\includegraphics[width=5.9in,height=\textheight]{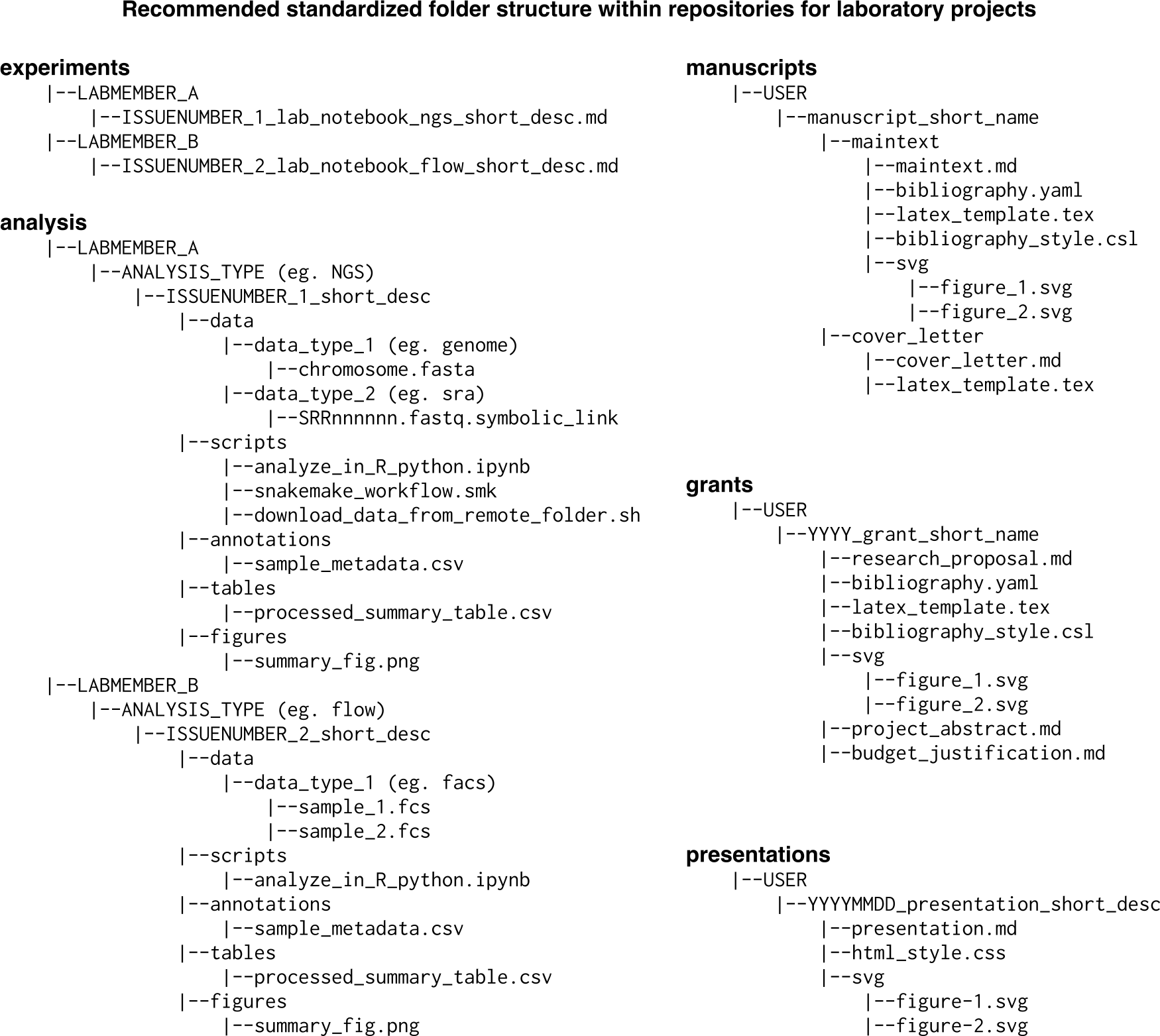}
\caption{\label{fig:folder_structure}}
\end{figure}

In our research group, we store all files relevant to project within a
single folder on our local computers. We use Git to track changes in
that folder, and synchronize it with a cloud-based GitHub repository. We
write documents in plain text with lightweight Markdown syntax, as often
as possible. Markdown enables focusing on content over formatting,
enables all changes to be tracked by Git , and can be easily converted
to other formats (PDF, DOCX, HTML) using open source software like
Pandoc. Within each repository, we use standardized subfolder names for
lab notebook entries, code, data, manuscripts, grants, and presentations
(Figure \ref{fig:folder_structure}). Within each of these subfolders,
each project contributor creates a separate folder to record their work,
even though every group member can contribute to all files in the
repository. Lab notebooks entries corresponding to distinct GitHub
issues are stored in separate files. We record all work pertinent to an
issue in lab notebook files, similar to traditional lab notebook
entries. Each lab notebook file includes the corresponding issue number
in its name and a link to the issue in its contents to enable easy
cross-referencing. All group members can access each other's lab
notebooks across different project repositories and participate in
discussion and troubleshooting steps by commenting on the corresponding
issue.

We also store data analysis materials within the GitHub project
repository, which ensures that experiment logs (lab notebooks) and
analysis scripts are tightly linked and easily referenced between
documents. Data analysis scripts, summary tables, and visualization
figures are stored in an `analysis' subfolder of the parent repository,
using the same hierarchical structure and naming convention as for lab
notebook entries. Separate folders are created for each issue with
subfolders for data, scripts, figures, tables, and sample metadata.
While we store small datasets in the GitHub repository as comma- or
tab-separated text files, larger datasets are stored either in private
Amazon Web Services S3 folders, or in public repositories such as the
Sequence Read Archive. We include short scripts to download data from
their long-term storage location, which also serves as a record of the
location of the data. A comma-delimited, sample metadata file is created
for every dataset, following tidy
principles\textsuperscript{\protect\hyperlink{ref-Wickham2014}{29}}, to
facilitate data analyses. Summary figures and tables are linked from the
lab notebook page as a record of how the data was analyzed, and are
linked from issue comments during data interpretation and
troubleshooting discussions.

In addition to lab notebooks and data analyses, we use the GitHub
repository to store manuscripts, grant proposals, and presentations
related to the project. We write manuscripts and grant proposals as
plain-text Markdown files, with one sentence per line, to enable easy
tracking of changes across Git commits. Markdown files can be easily
converted to DOCX, TeX, or PDF formats using Pandoc, which also provides
a suite of useful features for scientific writing such as citation
processing and template-based formatting to meet journal and funding
agency requirements. Multiple project contributors often edit manuscript
and grant proposal files in parallel, which can be readily combined
using the native merge functionality of Git while preserving the full
history of contributions. We prepare figures and presentation slides in
the widely used text-based scalable vector graphics (SVG) format, using
the powerful open-source software Inkscape. SVG files are rendered in
GitHub Markdown files and web browsers, and can be processed by most
commercial graphics design software. Presentations are written as
Markdown files with SVG-based images and speaker notes, which can then
be converted to slides in PPTX, PDF, or HTML formats using Pandoc.

In summary, the version control functionality of Git and GitHub that is
widely used to track changes to software code, can also be readily used
to track the work performed over the course of a project in a research
laboratory, including molecular biology ``wet'' labs. Git repositories
are drop-in replacements for lab notebooks while also providing a
tightly integrated structure for data analyses, manuscript writing,
grant preparation, and slide presentation tasks. By serving as a
centralized location for all project-related materials, they enable
contributors to reproduce and build on each other's work. Crucially,
even though we utilize GitHub for syncing repositories, project
materials themselves are independent of GitHub or any other platform,
which secures their long-term accessibility. Further, by providing a
chronological and transparent audit of all project-related
contributions, Git repositories incentivize collaboration between group
members, and unambiguously indicate individual contributions during
manuscript preparation and publication.

\hypertarget{use-containers-for-coding-and-writing-tasks}{%
\subsection{Use containers for coding and writing
tasks}\label{use-containers-for-coding-and-writing-tasks}}

Reproducibility and collaboration within a group critically depend on
the ability of members to run each other's data analyses and obtain the
same results. This allows members to build on each other's work,
troubleshoot issues, and reproduce results for manuscript writing and
grant preparation. However, software environments are difficult to
replicate, especially when they involve multiple programming languages,
packages, and dependencies, as is common for complex data analysis
workflows in molecular biology. Typical challenges that group members
and future collaborators face when attempting to rerun months- or
years-old analysis workflows include deprecated syntax, incompatible
package versions, and broken dependencies. Additionally, collaborative
writing tasks such as manuscript or grant preparation require a specific
set of software tools to produce the final document, which can be
difficult to replicate across different operating systems and software
versions. The time and effort required to troubleshoot software
incompatibility issues can be substantial, and can lead to the
abandonment of the task altogether.

The problem of replicating data analysis workflows and software
environments is a common challenge in software development as well as in
laboratory research. Computational researchers have long recognized this
problem and have relied on tools such as the Conda package
manager\textsuperscript{\protect\hyperlink{ref-Gruning2018a}{30}}, the
Snakemake workflow
manager\textsuperscript{\protect\hyperlink{ref-Koster2012}{31}}, and
Docker containers to address
it\textsuperscript{\protect\hyperlink{ref-Gruning2018}{6}}. Containers
are encapsulated, self-sufficient units that contain all the software
needed to run an analysis, and can be shared and run on any computer
that supports the container runtime environment. Public container
registries, like Docker Hub or
Biocontainers\textsuperscript{\protect\hyperlink{ref-DaVeigaLeprevost2017}{32}},
provide reproducibly-created containers, which can be used in data
analysis workflows without the need to install any software. However,
laboratory research groups have been slow to adopt these software
reproducibility tools, which are often perceived as too complex or
time-consuming to learn and use.

\begin{figure}
\centering
\includegraphics[width=5.1in,height=\textheight]{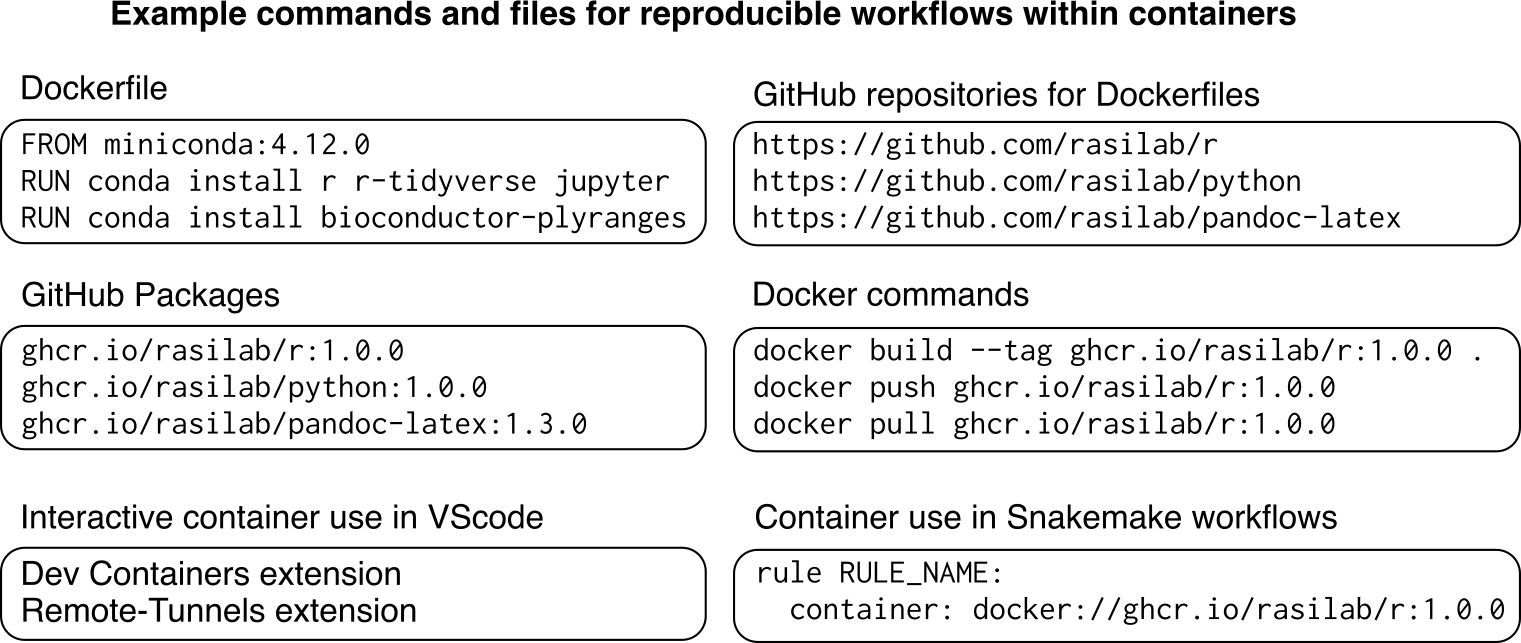}
\caption{\label{fig:containers}}
\end{figure}

In our research group, we use software containers to perform all data
analyses and writing tasks in reproducible software environments. We
take advantage of the Packages feature of GitHub to host our containers
in a centralized location
(\url{https://github.com/orgs/rasilab/packages}) that is free to use and
publicly accessible. Each container in our group's GitHub Packages
collection is linked to a dedicated GitHub repository to store the plain
text recipe, called a Dockerfile, for creating that container (Figure
\ref{fig:containers}). We have created a few general purpose containers
with R, Python, and Pandoc software that we routinely use for data
analysis and writing tasks in our group. Occasionally, we also create
new containers from scratch, or modify an existing container from a
public container registry to include a specific software package that is
needed for a specialized analysis. We use semantic versioning to tag
each container and its associated GitHub repository, which allows us to
unambiguously identify its contents and use the same container in our
data analysis workflows.

Our group uses containers in several ways for interactive data analyses,
writing tasks, and complex bioinformatic workflows (Figure
\ref{fig:containers}). Containers in our group's GitHub Packages can
also be used by external collaborators and readers of our published
manuscripts to reproduce data analyses. For interactive analyses on a
local computer with the Docker runtime environment, one of the general
purpose containers from our group's GitHub Packages registry can be
copied (`pulled') to the local computer. Then the user can either access
the container through the Remote-Containers extension in the Visual
Studio Code editor, or run the container in a terminal window. Running
containers can be used to convert Markdown files to other formats using
Pandoc, or to run R or Python scripts for data analyses. In shared
computing environments such as high-performance computing clusters,
containers can be downloaded to a shared location and run using the
Apptainer (Singularity) runtime environment. Apptainer containers in
remote computing environments can be used in workflow management tools,
like Snakemake, to run multi-step computational analyses, or accessed
from a personal computer using the Remote-Tunnels extension in the
Visual Studio Code editor for interactive analyses.

In summary, containers, which are widely used in software development
and computational research, facilitate reproducible and collaborative
data analysis and writing within laboratory research groups. GitHub
Packages provide a centralized location for groups to store their
frequently used software environments as Docker containers, and share
them within the group and with the external scientific community. Once
containers are set up and optimized for a group's common workflows by an
experienced group member or a bioinformatician, they can be used by all
group members, far into the future, without change or detailed know-how.
In our experience, containers are particularly useful for new members to
get up to speed on our group's data analysis and writing workflows
without struggling to replicate the necessary software environments.

\hypertarget{discussion}{%
\subsection{Discussion}\label{discussion}}

In this practical guide, we have described our group's approach for
tracking all stages of laboratory research from idea generation and
experimental design, to data analysis and manuscript writing. Our
approach is motivated by the recognition that established software
development practices provide a concrete framework for addressing the
reproducibility and collaboration challenges faced by laboratory
research groups. We have adopted widely used features from software
development workflows, such as issues, version control, and containers,
and adapted them to the specific needs of a molecular biology
laboratory. We have illustrated our approach using the GitHub platform,
but other platforms such as GitLab and Bitbucket offer similar
functionalities.

We recognize that adopting the approach outlined here can involve a
steep learning curve, especially for laboratory research groups with
limited computational experience. However, there are several benefits to
using GitHub for laboratory research that we believe outweigh the
initial investment of time and effort. First, Git and GitHub are widely
used in both academia and industry, and thus the organization and
documentation practices we describe are highly transferrable skills for
trainees. Second, Git and GitHub have comprehensive and user-friendly
documentation (Table \protect\hyperlink{table-1}{1}), and a number of
tutorials and forums are available online to help new users troubleshoot
any issues that arise. Furthermore, these tools are so widely used that
virtually any bioinformatician or bioinformatic core at an institution
can help new teams set up and troubleshoot their GitHub workflow. Third,
the workflow and features described here are highly modular. Therefore,
teams can incrementally adopt them, while still deriving benefits to
their overall research productivity. Finally, the approach described
here costs nothing to implement, and can be used by any research group
regardless of their size, funding level, or institutional affiliation.

While this guide covers the core functionalities of Git and GitHub for
laboratory research, there are additional features that can further
enhance collaboration and productivity. For instance, Git branching
allows finer control over collaborative data analysis and writing across
large teams, by allowing parallel development of different threads of
ideas while retaining the history. GitHub Actions can enable the
creation of automated workflows for repetitive tasks, such as updating
lab website and documentation when changes are pushed to a GitHub
repository. Cloud-based containers, such as GitHub Codespaces, can
enable groups to perform most of their analysis and writing tasks from
within a web browser without the need to install any software on their
local computer. Wiki and Discussions features in GitHub allow
documentation of protocols and open-ended conversations that are outside
the scope of specific projects. These features, while beyond the scope
of this introductory guide, can be adopted by laboratory research groups
as they become more comfortable with Git and GitHub.

The organizational approach described here is tailored to the lifecycle
of a conventional laboratory research project from idea generation to
manuscript writing. Nevertheless, this workflow offers rich
possibilities for a more reproducible and collaborative research
enterprise at the institutional and community levels. For instance,
GitHub issues can be used after manuscript publication to handle reagent
requests and answer protocol-related questions, thus providing a
centralized location for community feedback and engagement. Institutions
can provide backup and support for GitHub repositories, thereby ensuring
that the research record is preserved even if the original research
group is no longer active or associated with the institution. With
public GitHub repositories, community experts can contribute ideas and
feedback during the research process, and their contributions will be
visible in the repository history and issue comments. The GitHub
repository itself can serve as a living manuscript, with GitHub releases
or tags constituting different versions of the manuscript as it evolves
over time. Thus, the approach outlined here could potentially accelerate
the pace of scientific discovery by enabling faster dissemination of
results and fostering more collaboration opportunities.

\hypertarget{author-contributions}{%
\section{Author Contributions}\label{author-contributions}}

K.Y.C, M.T.M., and A.R.S. wrote the manuscript. A.R.S. acquired funding.

\hypertarget{acknowledgements}{%
\section{Acknowledgements}\label{acknowledgements}}

We thank members of the Subramaniam lab, the Basic Sciences Division,
and the Computational Biology Program at Fred Hutch for discussions, and
Rechel Geiger, Pravrutha Raman, and Jamie Yelland for feedback on the
manuscript. This research was funded by NIH R35 GM119835 (A.R.S.), NSF
MCB 1846521 (A.R.S.), NIH R01 AT012826 (A.R.S.), and the Hanna H. Gray
Fellowship GT16007 (M.T.M). The funders had no role in decision to
publish or preparation of the manuscript.

\hypertarget{competing-interests}{%
\section{Competing interests}\label{competing-interests}}

None

\nolinenumbers

\hypertarget{references}{%
\section{References}\label{references}}

\hypertarget{refs}{}
\begin{CSLReferences}{0}{0}
\leavevmode\vadjust pre{\hypertarget{ref-Hrynaszkiewicz2020}{}}%
\CSLLeftMargin{1. }
\CSLRightInline{Hrynaszkiewicz, I., Simons, N., Hussain, A., Grant, R.
\& Goudie, S. \href{https://doi.org/10.5334/dsj-2020-005}{Developing a
Research Data Policy Framework for All Journals and Publishers}.
\emph{Data Science Journal} \textbf{19}, 5--5 (2020).}

\leavevmode\vadjust pre{\hypertarget{ref-Tedersoo2021}{}}%
\CSLLeftMargin{2. }
\CSLRightInline{Tedersoo, L. \emph{et al.}
\href{https://doi.org/10.1038/s41597-021-00981-0}{Data sharing practices
and data availability upon request differ across scientific
disciplines}. \emph{Sci Data} \textbf{8}, 192 (2021).}

\leavevmode\vadjust pre{\hypertarget{ref-nih2023}{}}%
\CSLLeftMargin{3. }
\CSLRightInline{Data Management and Sharing Policy \textbar{} Data
Sharing.
\url{https://sharing.nih.gov/data-management-and-sharing-policy}.}

\leavevmode\vadjust pre{\hypertarget{ref-Heil2021}{}}%
\CSLLeftMargin{4. }
\CSLRightInline{Heil, B. J. \emph{et al.}
\href{https://doi.org/10.1038/s41592-021-01256-7}{Reproducibility
standards for machine learning in the life sciences}. \emph{Nat Methods}
\textbf{18}, 1132--1135 (2021).}

\leavevmode\vadjust pre{\hypertarget{ref-Noble2009}{}}%
\CSLLeftMargin{5. }
\CSLRightInline{Noble, W. S.
\href{https://doi.org/10.1371/journal.pcbi.1000424}{A Quick Guide to
Organizing Computational Biology Projects}. \emph{PLOS Computational
Biology} \textbf{5}, e1000424 (2009).}

\leavevmode\vadjust pre{\hypertarget{ref-Gruning2018}{}}%
\CSLLeftMargin{6. }
\CSLRightInline{Grüning, B. \emph{et al.}
\href{https://doi.org/10.1038/s41592-018-0046-7}{Bioconda: sustainable
and comprehensive software distribution for the life sciences}.
\emph{Nat Methods} \textbf{15}, 475--476 (2018).}

\leavevmode\vadjust pre{\hypertarget{ref-Jenkins2023}{}}%
\CSLLeftMargin{7. }
\CSLRightInline{Jenkins, G. B. \emph{et al.}
\href{https://doi.org/10.1002/ece3.9961}{Reproducibility in ecology and
evolution: Minimum standards for data and code}. \emph{Ecol Evol}
\textbf{13}, e9961 (2023).}

\leavevmode\vadjust pre{\hypertarget{ref-Diaba-Nuhoho2021}{}}%
\CSLLeftMargin{8. }
\CSLRightInline{Diaba-Nuhoho, P. \& Amponsah-Offeh, M.
\href{https://doi.org/10.1186/s13104-021-05875-3}{Reproducibility and
research integrity: the role of scientists and institutions}. \emph{BMC
Research Notes} \textbf{14}, 451 (2021).}

\leavevmode\vadjust pre{\hypertarget{ref-Baker2016}{}}%
\CSLLeftMargin{9. }
\CSLRightInline{Baker, M. \href{https://doi.org/10.1038/533452a}{1,500
scientists lift the lid on reproducibility}. \emph{Nature} \textbf{533},
452--454 (2016).}

\leavevmode\vadjust pre{\hypertarget{ref-Ram2013}{}}%
\CSLLeftMargin{10. }
\CSLRightInline{Ram, K. \href{https://doi.org/10.1186/1751-0473-8-7}{Git
can facilitate greater reproducibility and increased transparency in
science}. \emph{Source Code for Biology and Medicine} \textbf{8}, 7
(2013).}

\leavevmode\vadjust pre{\hypertarget{ref-Lowndes2017}{}}%
\CSLLeftMargin{11. }
\CSLRightInline{Lowndes, J. S. S. \emph{et al.}
\href{https://doi.org/10.1038/s41559-017-0160}{Our path to better
science in less time using open data science tools}. \emph{Nat Ecol
Evol} \textbf{1}, 1--7 (2017).}

\leavevmode\vadjust pre{\hypertarget{ref-Braga2023}{}}%
\CSLLeftMargin{12. }
\CSLRightInline{Braga, P. H. P. \emph{et al.}
\href{https://doi.org/10.1111/2041-210X.14108}{Not just for programmers:
How GitHub can accelerate collaborative and reproducible research in
ecology and evolution}. \emph{Methods in Ecology and Evolution}
\textbf{14}, 1364--1380 (2023).}

\leavevmode\vadjust pre{\hypertarget{ref-Berg2017}{}}%
\CSLLeftMargin{13. }
\CSLRightInline{Berg, J.
\href{https://doi.org/10.1126/science.aar2406}{Editorial retraction}.
\emph{Science} \textbf{358}, 458--458 (2017).}

\leavevmode\vadjust pre{\hypertarget{ref-Wosen2023}{}}%
\CSLLeftMargin{14. }
\CSLRightInline{Wosen, J. Genentech review of Tessier-Lavigne paper
finds no evidence of fraud --- but hints at a different misconduct case.
\emph{STAT}
\url{https://www.statnews.com/2023/04/06/genentech-marc-tessier-lavigne-stanford-misconduct-investigation/}
(2023).}

\leavevmode\vadjust pre{\hypertarget{ref-Ryan2010}{}}%
\CSLLeftMargin{15. }
\CSLRightInline{Ryan, P. Keeping a Lab Notebook: Basic Principles and
Best Practices. \emph{Office of Intramural Training and Education,
National Institutes of Health} (2010).}

\leavevmode\vadjust pre{\hypertarget{ref-Higgins2022}{}}%
\CSLLeftMargin{16. }
\CSLRightInline{Higgins, S. G., Nogiwa-Valdez, A. A. \& Stevens, M. M.
\href{https://doi.org/10.1038/s41596-021-00645-8}{Considerations for
implementing electronic laboratory notebooks in an academic research
environment}. \emph{Nat Protoc} \textbf{17}, 179--189 (2022).}

\leavevmode\vadjust pre{\hypertarget{ref-Johnson2003}{}}%
\CSLLeftMargin{17. }
\CSLRightInline{Johnson, J. N. \& Dubois, P. F.
\href{https://doi.org/10.1109/MCISE.2003.1238707}{Issue Tracking}.
\emph{Computing in Science and Engg.} \textbf{5}, 71--77 (2003).}

\leavevmode\vadjust pre{\hypertarget{ref-Blischak2016}{}}%
\CSLLeftMargin{18. }
\CSLRightInline{Blischak, J. D., Davenport, E. R. \& Wilson, G.
\href{https://doi.org/10.1371/journal.pcbi.1004668}{A Quick Introduction
to Version Control with Git and GitHub}. \emph{PLOS Computational
Biology} \textbf{12}, e1004668 (2016).}

\leavevmode\vadjust pre{\hypertarget{ref-Moreau2023}{}}%
\CSLLeftMargin{19. }
\CSLRightInline{Moreau, D., Wiebels, K. \& Boettiger, C.
\href{https://doi.org/10.1038/s43586-023-00236-9}{Containers for
computational reproducibility}. \emph{Nat Rev Methods Primers}
\textbf{3}, 1--16 (2023).}

\leavevmode\vadjust pre{\hypertarget{ref-2023}{}}%
\CSLLeftMargin{20. }
\CSLRightInline{Key GitHub Statistics in 2024 (Users, Employees, and
Trends). \emph{Kinsta®} \url{https://kinsta.com/blog/github-statistics/}
(2023).}

\leavevmode\vadjust pre{\hypertarget{ref-stackoverflow2023}{}}%
\CSLLeftMargin{21. }
\CSLRightInline{Stack Overflow Developer Survey 2022. \emph{Stack
Overflow} \url{https://survey.stackoverflow.co/2022}.}

\leavevmode\vadjust pre{\hypertarget{ref-octoverse2022}{}}%
\CSLLeftMargin{22. }
\CSLRightInline{The state of open source software. \emph{The State of
the Octoverse}
\url{https://octoverse.github.com/2022/state-of-open-source}.}

\leavevmode\vadjust pre{\hypertarget{ref-Cadwallader2022}{}}%
\CSLLeftMargin{23. }
\CSLRightInline{Cadwallader, L. \& Hrynaszkiewicz, I.
\href{https://doi.org/10.7717/peerj.13933}{A survey of researchers' code
sharing and code reuse practices, and assessment of interactive notebook
prototypes}. \emph{PeerJ} \textbf{10}, e13933 (2022).}

\leavevmode\vadjust pre{\hypertarget{ref-Perkel2016}{}}%
\CSLLeftMargin{24. }
\CSLRightInline{Perkel, J.
\href{https://doi.org/10.1038/538127a}{Democratic databases: science on
GitHub}. \emph{Nature} \textbf{538}, 127--128 (2016).}

\leavevmode\vadjust pre{\hypertarget{ref-Perez-Riverol2016}{}}%
\CSLLeftMargin{25. }
\CSLRightInline{Perez-Riverol, Y. \emph{et al.}
\href{https://doi.org/10.1371/journal.pcbi.1004947}{Ten Simple Rules for
Taking Advantage of Git and GitHub}. \emph{PLoS Comput Biol}
\textbf{12}, e1004947 (2016).}

\leavevmode\vadjust pre{\hypertarget{ref-Stanisic2015}{}}%
\CSLLeftMargin{26. }
\CSLRightInline{Stanisic, L., Legrand, A. \& Danjean, V.
\href{https://doi.org/10.1145/2723872.2723881}{An Effective Git And
Org-Mode Based Workflow For Reproducible Research}. \emph{Operating
Systems Review} \textbf{49}, 61 (2015).}

\leavevmode\vadjust pre{\hypertarget{ref-Chure2022}{}}%
\CSLLeftMargin{27. }
\CSLRightInline{Chure, G. Be Prospective, Not Retrospective: A
Philosophy for Advancing Reproducibility in Modern Biological Research.
(2022)
doi:\href{https://doi.org/10.48550/arXiv.2210.02593}{10.48550/arXiv.2210.02593}.}

\leavevmode\vadjust pre{\hypertarget{ref-Himmelstein2019}{}}%
\CSLLeftMargin{28. }
\CSLRightInline{Himmelstein, D. S. \emph{et al.}
\href{https://doi.org/10.1371/journal.pcbi.1007128}{Open collaborative
writing with Manubot}. \emph{PLOS Computational Biology} \textbf{15},
e1007128 (2019).}

\leavevmode\vadjust pre{\hypertarget{ref-Wickham2014}{}}%
\CSLLeftMargin{29. }
\CSLRightInline{Wickham, H.
\href{https://doi.org/10.18637/jss.v059.i10}{Tidy Data}. \emph{J. Stat.
Soft.} \textbf{59}, (2014).}

\leavevmode\vadjust pre{\hypertarget{ref-Gruning2018a}{}}%
\CSLLeftMargin{30. }
\CSLRightInline{Grüning, B. \emph{et al.}
\href{https://doi.org/10.1016/j.cels.2018.03.014}{Practical
Computational Reproducibility in the Life Sciences}. \emph{Cell Syst}
\textbf{6}, 631--635 (2018).}

\leavevmode\vadjust pre{\hypertarget{ref-Koster2012}{}}%
\CSLLeftMargin{31. }
\CSLRightInline{Köster, J. \& Rahmann, S.
\href{https://doi.org/10.1093/bioinformatics/bts480}{Snakemake---a
scalable bioinformatics workflow engine}. \emph{Bioinformatics}
\textbf{28}, 2520--2522 (2012).}

\leavevmode\vadjust pre{\hypertarget{ref-DaVeigaLeprevost2017}{}}%
\CSLLeftMargin{32. }
\CSLRightInline{Da Veiga Leprevost, F. \emph{et al.}
\href{https://doi.org/10.1093/bioinformatics/btx192}{BioContainers: an
open-source and community-driven framework for software
standardization}. \emph{Bioinformatics} \textbf{33}, 2580--2582 (2017).}

\end{CSLReferences}

\end{document}